\documentclass[conference]{IEEEtran}
\IEEEoverridecommandlockouts
\usepackage{makecell}
\usepackage[dvips]{graphicx}
\usepackage{latexsym}
\usepackage{amssymb}
\usepackage{amsmath}
\usepackage{bm}
\usepackage{multirow}
\usepackage{xcolor}
\usepackage{lipsum}
\usepackage{multicol}
\usepackage{enumerate}
\usepackage{algorithm}
\usepackage{algorithmicx}
\usepackage{algpseudocode}
\usepackage{amsmath}
\usepackage{graphicx}
\usepackage{subfig}
\usepackage[utf8]{inputenc}
\usepackage{float}

\def\BibTeX{{\rm B\kern-.05em{\sc i\kern-.025em b}\kern-.08em
    T\kern-.1667em\lower.7ex\hbox{E}\kern-.125emX}}
\begin{document}

\title{Block-Level Interference Exploitation Precoding without Symbol-by-Symbol Optimization
\thanks{This work is supported in part by the National Natural Science Foundation of China under Grant 62101422, in part by the Science and Technology Program of Shaanxi Province under Grant 2021KWZ-01, and in part by the Fundamental Research Funds for the Central Universities under Grant xzy012020007.}
}
\author{\IEEEauthorblockN{Ang Li$^*$, Chao Shen$^\ddag$, Xuewen Liao$^*$, Christos Masouros$^\dag$, and A. Lee Swindlehurst$^\P$}
\IEEEauthorblockA{
School of Information and Communications Engineering, Xi'an Jiaotong University, Xi'an, China$^*$\\
Shenzhen Research Institute of Big Data, Shenzhen, China$^\ddag$\\
Department of Electronic and Electrical Engineering, University College London, London, UK$^\dag$\\
Center for Pervasive Communications and Computing, University of California, Irvine, USA$^\P$\\
Email: \{ang.li.2020, yeplos\}@xjtu.edu.cn$^*$, chaoshen@sribd.cn$^\ddag$, c.masouros@ucl.ac.uk$^\dag$, swindle@uci.edu$^\P$}
}

\maketitle

\begin{abstract}
Symbol-level precoding (SLP) based on the concept of constructive interference (CI) is shown to be superior to traditional block-level precoding (BLP), however at the cost of a symbol-by-symbol optimization during the precoding design. In this paper, we propose a CI-based block-level precoding (CI-BLP) scheme for the downlink transmission of a multi-user multiple-input single-output (MU-MISO) communication system, where we design a constant precoding matrix to a block of symbol slots to exploit CI for each symbol slot simultaneously. A single optimization problem is formulated to maximize the minimum CI effect over the entire block, thus reducing the computational cost of traditional SLP as the optimization problem only needs to be solved once per block. By leveraging the Karush-Kuhn-Tucker (KKT) conditions and the dual problem formulation, the original optimization problem is finally shown to be equivalent to a quadratic programming (QP) over a simplex. Numerical results validate our derivations and exhibit superior performance for the proposed CI-BLP scheme over traditional BLP and SLP methods, thanks to the relaxed block-level power constraint.
\end{abstract}

\begin{IEEEkeywords}
MIMO, symbol-level precoding, constructive interference, interference exploitation, optimization.
\end{IEEEkeywords}

\section{Introduction}
Interference management plays a crucial role for reliable communication in multiple-input multiple-output (MIMO) systems. In the downlink transmission of a multi-user MIMO system, precoding is essential for realizing spatial multiplexing, and a number of block-level precoding (BLP) approaches have been designed in the literature to manage multi-user interference. This includes zero-forcing (ZF) precoding \cite{r5} and block-diagonalization (BD) precoding \cite{r_BD}, as well as the regularized ZF (RZF) precoding that enhances the performance of ZF precoding \cite{r6}. Furthermore, optimization-based precoding schemes have been proposed in the literature for additional performance improvements over closed-form precoders, for example the signal-to-interference-plus-noise ratio (SINR) balancing precoding \cite{r8}, \cite{r9}, and the weighted minimum mean-squared error (W-MMSE) precoder \cite{WMMSE}.

More recently, the concept of constructive interference (CI) has been introduced to the precoder design in MIMO communications \cite{r29}, \cite{r30}, where it is shown that by further exploiting the data symbol information in addition to the channel state information (CSI), instantaneous interference existing in multi-user transmission can be categorized into constructive and destructive, and judicious precoding approaches have been designed to exploit the constructive part of multi-user interference and meanwhile transform the destructive part into constructive, leading to significant performance improvements \cite{r19}, \cite{r31}. However, it should be mentioned that the performance benefits of CI-based precoding come at the cost of a symbol-by-symbol design methodology, i.e., symbol-level precoding (SLP) where the precoder must be designed for each symbol slot is required. This poses a significant computational burden on the multi-user MIMO communication system, because the base station (BS) needs to solve an independent optimization problem for each symbol slot. To alleviate the computational costs, several studies attempt to reduce the complexity of the CI-SLP optimization problem, including derivations of the optimal precoding structure for CI-SLP with efficient iterative algorithms \cite{r31}, sub-optimal solutions \cite{r33}, and deep learning-based methods \cite{DL-1}. Despite the above attempts, all the above approaches still require solving an optimization problem at the symbol level, i.e., the total number of CI-SLP problems that must be solved in a channel coherence interval is however not reduced.

In this paper, for the first time in the literature we propose a CI-based block-level precoding (CI-BLP) approach that applies a constant precoding matrix to a block of symbol slots in a downlink multi-user multiple-input single-output (MU-MISO) system, where CI is achieved for all the symbol slots of the transmission block simultaneously. Based on the `symbol-scaling' CI metric, a single optimization problem is formulated to maximize the minimum CI effect over all symbol slots subject to a block- rather than symbol-level power budget. By leveraging the Lagrangian method and studying the corresponding dual problem, the original CI-BLP optimization problem is finally shown to be equivalent to a QP optimization over a simplex. Numerical results demonstrate that the proposed CI-BLP approach offers an improved error-rate performance compared with traditional CI-SLP approaches thanks to the relaxed block-level power constraint, which meanwhile require fewer computational costs as the optimization problem only needs to be solved once per block.

\section{System Model and Constructive Interference}
\subsection{System Model}
The downlink transmission of a MU-MISO communication system is considered, where a total number of $K$ single-antenna users are served by a BS with $N_\text{T}$ transmit antennas, and $K \le N_\text{T}$. We focus on the transmission of a block of symbol slots, where we introduce ${\bf s}^n = \left [ {s_1^n, s_2^n, \cdots, s_K^n} \right]^\text{T} \in {\mathbb C}^{K \times 1}$ as the data symbol vector in the $n$-th slot, drawn from normalized PSK constellations. Accordingly, the received signal for user $k$ in the $n$-th symbol slot can be expressed as
\begin{equation}
y_k^n= {\bf h}_k^\text{T}{\bf W}{\bf s}^n + z_k^n,
\label{eq_1}
\end{equation}
where ${\bf h}_k \in {\mathbb C}^{N_\text{T} \times 1}$ represents the channel vector between the transmit antenna array and the $k$-th user\footnote{Since we focus on deriving the optimal precoding structure for the proposed CI-BLP method, perfect CSI is assumed throughout the paper.}, which is constant within the considered block, and $z_k^n$ is additive Gaussian noise with zero mean and variance $\sigma^2$. ${\bf W} \in {\mathbb C}^{N_\text{T} \times K}$ is the precoding matrix that applies to all ${\bf s}^n$ in the block.

\begin{figure}[!t]
\centering
\includegraphics[scale=0.3]{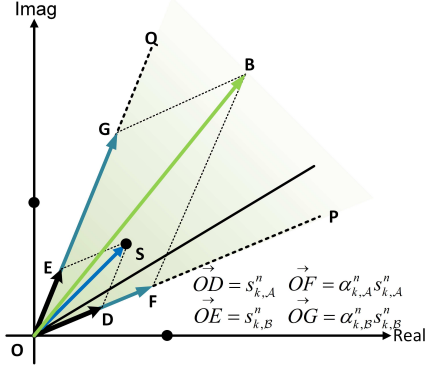}
\caption{An illustration for `symbol-scaling' CI metric, 8PSK}
\end{figure}

\subsection{Constructive Interference}
CI is the interference that is able to push the signals of interest further away from the decision boundaries of their modulated symbol constellation, such that the received signals have a higher probability to be correctly detected \cite{r29}. In this paper, we employ the `symbol-scaling' CI metric for PSK modulation to ease our subsequent derivations \cite{r32}. To illustrate the concept of the `symbol-scaling' CI metric, in Fig. 1 we depict one quarter of a 8PSK constellation as an example, where without loss of generality we denote $\vec {OS}=s_k^n$ as the data symbol of interest for user $k$ in the $n$-th symbol slot, and $\vec{OB}={\bf h}_k^\text{T}{\bf W}{\bf s}^n$ as the corresponding received signal excluding noise. The `symbol-scaling' CI metric decomposes the data symbols and the received signals along their corresponding decision boundaries, i.e., $\vec {OS}=\vec{OD} + \vec{OE}$ and $\vec {OB}=\vec{OF} + \vec{OG}$. To obtain a better error-rate performance, CI precoding aims to push the received signal $\vec{OB}$ further away from the decision boundaries ($\vec{OP}$ and $\vec{OQ}$ in Fig. 1), and this is equivalent to increasing the amplitude for $\vec{OF}$ and $\vec{OG}$ as much as possible, as will be shown in Section III-A mathematically.

\section{Proposed Block-Level Interference Exploitation Precoding}
In this section, the proposed CI-BLP approach is introduced. As shown in Fig. 2, the proposed CI-BLP aims to exploit CI for each symbol slot simultaneously with a constant precoding matrix $\bf W$ that is applied to all symbol slots within the block, where $N$ is the length of the considered transmission block. In the following, we construct the optimization problem for the proposed CI-BLP method and derives its optimal precoding structure.
%Compared with traditional CI-SLP, our proposed CI-SLP removes the necessity of symbol-by-symbol optimization in the precoding design, thus greatly reducing the computational complexity as the optimization problem only needs to be solved once per block. On the other hand, when compared to traditional BLP, our proposed CI-BLP further exploits the information of the data symbols available at the BS for further performance improvements, i.e., conventional BLP is data-independent while CI-BLP is data-dependent.

\begin{figure}[!t]
\centering
\includegraphics[scale=0.8]{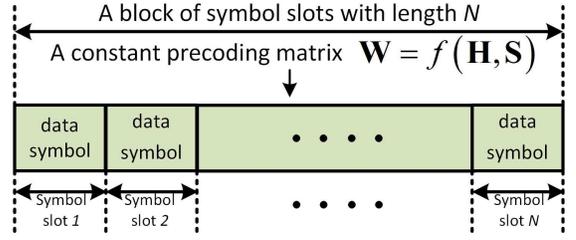}
\caption{The proposed CI-BLP methodology}
\end{figure}

\newcounter{mytempeqncnt1}
\begin{figure*}[!t]
\normalsize
\setcounter{equation}{13}

\begin{equation}
\begin{aligned}
&{\cal L}\left ( {\bf \hat W} , {t}, {\delta_{k}^{n}}, {\mu}  \right ) \\
& = -{t} +\sum_{n=1}^{N} \sum_{k=1}^{2K} {\delta_k^n}\left [ {{t}- \left( {{\bf a}_k^n} \right)^\text{T}{\bf \hat W} {\bf s}_\text{E}^n - \left( {{\bf b}_k^n} \right)^\text{T}{\bf \hat W} {\bf c}_\text{E}^n} \right ] + {\mu} \left[{\sum_{n=1}^{N} \left ( {\bf s}_\text{E}^n \right )^\text{T} \left ( {{\bf P} {\bf \hat W} + {\bf Q} {\bf \hat W} {\bf T}} \right )^\text{T}\left ( {{\bf P} {\bf \hat W} + {\bf Q} {\bf \hat W} {\bf T}} \right ) {\bf s}_\text{E}^n -N p_0 }\right] \\
&=\left ( {\sum_{n=1}^{N} {\bf 1}^\text{T}{\bm{\delta}}^n -1 } \right ) {t} -\sum_{n=1}^{N}\left( {{\bm{ \delta}}^n} \right)^\text{T}{\bf A}^n{\bf \hat W}{\bf s}_\text{E}^n -\sum_{n=1}^{N}\left( {{\bm{\delta}}^n} \right)^\text{T}{\bf B}^n{\bf \hat W}{\bf c}_\text{E}^n + {\mu}\sum_{n=1}^{N} \left ( {\bf s}_\text{E}^n  \right )^\text{T} {\bf \hat W}^\text{T} {\bf \hat W} {\bf s}_\text{E}^n + {\mu}\sum_{n=1}^{N} \left ( {\bf c}_\text{E}^n  \right )^\text{T} {\bf \hat W}^\text{T} {\bf \hat W} {\bf c}_\text{E}^n -{u_0}Np_0
\end{aligned}
\label{eq_14}
\end{equation}

\hrulefill
\vspace*{4pt}
\end{figure*}

\subsection{Problem Formulation}
Following the principle of the `symbol-scaling' CI metric in Fig. 1, the data symbol $\vec{OS}=s_k^n$ for user $k$ in the $n$-th symbol slot and the corresponding received signal excluding noise $\vec{OB}={\bf h}_k^\text{T}{\bf W}{\bf s}^n$ can be decomposed into:
\setcounter{equation}{1}
\begin{equation}
\begin{aligned}
 \vec{OS}&=\vec{OD}+\vec{OE} {\kern 1pt} \Rightarrow {\kern 1pt} s_k^n = s_{k, \cal A}^n +  s_{k, \cal B}^n, \\
 \vec{OB}&=\vec{OF}+\vec{OG} {\kern 2pt} \Rightarrow {\kern 1pt} {\bf h}_k^\text{T}{\bf W}{\bf s}^n =\alpha_{k, \cal A}^n s_{k, \cal A}^n + \alpha_{k, \cal B}^n s_{k, \cal B}^n,
\end{aligned}
\label{eq_2}
\end{equation}
where $\alpha_{k, \cal A}^n \ge 0$ and $\alpha_{k, \cal B}^n \ge 0$ represents the scaling effect. We define ${\bm \alpha}_\text{E}^n \in {\mathbb R}^{2K \time 1}$ as
\begin{equation}
{\bm \alpha}_\text{E}^n=\left[ {\alpha_{1, \cal A}^n, \alpha_{2, \cal A}^n, \cdots, \alpha_{K, \cal A}^n, \alpha_{1, \cal B}^n, \alpha_{2, \cal B}^n, \cdots, \alpha_{K, \cal B}^n} \right] ^\text{T},
\label{eq_3}
\end{equation}
and by following the transformations in \cite{r25}, ${\bm \alpha}_\text{E}^n$ can be further expressed as
\begin{equation}
{\bm \alpha}_\text{E}^n={\bf M}^n {\bf W}_\text{E} {\bf s}_\text{E}^n,
\label{eq_4}
\end{equation}
where the construction of ${\bf M}^n \in {\mathbb R}^{2K \times 2N_\text{T}}$ directly follows Section IV-A of \cite{r25}, which is shown in the appendix of \cite{r39}. ${\bf W}_\text{E} \in {\mathbb R}^{2N_\text{T} \times 2K}$ and ${\bf s}_\text{E}^n \in {\mathbb R}^{2K \times 1}$ in \eqref{eq_4} are defined as
\begin{equation}
{\bf W}_\text{E} = \left [ \begin{matrix}
 \Re \left ( \bf W \right )  & - \Im \left ( \bf W \right ) \\
 - \Im \left ( \bf W \right ) & \Re \left ( \bf W \right )
\end{matrix} \right ], {\kern 3pt} {\bf s}_\text{E}^n= \left[ {\Re \left( {{\bf s}^n} \right)^\text{T}, \Im \left( {{\bf s}^n} \right)^\text{T} } \right]^\text{T}.
\label{eq_5}
\end{equation}

From Fig. 1, we observe that a larger value of $\alpha_{k, \cal A}^n$ or $\alpha_{k, \cal B}^n$ represents a larger distance to the decision boundaries, leading to a better error-rate performance. Accordingly, the proposed CI-BLP approach aims to maximize the minimum entry in ${\bm \alpha}_\text{E}^n$ for all symbol slots within the block, and the corresponding optimization problem can be constructed as:
\begin{equation}
\begin{aligned}
&\mathcal{P}_0: {\kern 3pt} \max_{{\bf W}_\text{E}} \min_{k,n} {\kern 3pt} \alpha_k^n \\
&{\kern 6pt} \text{s.t.} {\kern 13pt} {\bf C1:} {\kern 3pt} {\bm \alpha}_\text{E}^n={\bf M}^n {\bf W}_\text{E} {\bf s}_\text{E}^n, {\kern 3pt} \forall n \le N, \\
&{\kern 31pt} {\bf C2:} {\kern 3pt} \sum_{n=1}^{N} \left \| {{\bf W}_\text{E} {\bf s}_\text{E}^n  }  \right \| _2^2 \le N{p_0},
\label{eq_6}
\end{aligned}
\end{equation}
where $\alpha_k^n$ represents the $k$-th entry in ${\bm \alpha}_\text{E}^n$, and $p_0$ represents the transmit power budget per symbol slot.

\newcounter{mytempeqncnt2}
\begin{figure*}[!t]
\normalsize
\setcounter{equation}{15}

\begin{IEEEeqnarray}{rCl}
\IEEEyesnumber
\frac{{\partial {\cal L}}}{{\partial t}} =\sum_{n=1}^{N} {\bf 1}^\text{T}{\bm{\delta}}^n -1 =0  {\kern 30pt} \IEEEyessubnumber* \label{eq_16a} \\
\frac{{\partial {\cal L}}}{{\partial {\bf \hat W}}} = -\sum_{n=1}^{N} \left [ {\left ( {{\bm{\delta}}^n} \right ) }^\text{T} {\bf A}^n \right ] ^\text{T} \left ( {\bf s}_\text{E}^n  \right )^\text{T} -\sum_{n=1}^{N} \left [ {\left ( {{\bm{\delta}}^n} \right ) }^\text{T} {\bf B}^n \right ] ^\text{T} \left ( {\bf c}_\text{E}^n  \right )^\text{T} + 2{u_0} {\bf \hat W}\left [ {\sum_{n=1}^{N} {\bf s}_\text{E}^n  \left ( {\bf s}_\text{E}^n \right )^\text{T} + \sum_{n=1}^{N} {\bf c}_\text{E}^n  \left ( {\bf c}_\text{E}^n \right )^\text{T} }  \right ]  = {\bf 0}  {\kern 30pt} \label{eq_16b} \\
{\delta_k^n}\left [ {{t}- \left( {{\bf a}_k^n} \right)^\text{T}{\bf \hat W} {\bf s}_\text{E}^n - \left( {{\bf b}_k^n} \right)^\text{T}{\bf \hat W} {\bf c}_\text{E}^n} \right ]=0, {\kern 3pt} {\delta_k^n} \ge 0, {\kern 3pt} \forall k \le 2K, {\kern 3pt} \forall n \le N {\kern 30pt} \label{eq_16c} \\
{\mu}\left [ \sum_{n=1}^{N} \left ( {\bf s}_\text{E}^n  \right )^\text{T} {\bf \hat W}^\text{T} {\bf \hat W} {\bf s}_\text{E}^n + \sum_{n=1}^{N} \left ( {\bf c}_\text{E}^n  \right )^\text{T} {\bf \hat W}^\text{T} {\bf \hat W} {\bf c}_\text{E}^n -Np_0 \right ] =0, {\kern 3pt} {\mu} \ge 0 {\kern 30pt} \label{eq_16d}
\end{IEEEeqnarray}

\hrulefill
\vspace*{4pt}
\end{figure*}

\newcounter{mytempeqncnt3}
\begin{figure*}[!t]
\normalsize
\setcounter{equation}{21}

\begin{equation}
\begin{aligned}
{\cal U} & = \max_{\left\{{{\bm{\delta}}^m}\right\}, {\mu}} -\sum_{m=1}^{N} \left ( {{\bm{\delta}}^m} \right )^\text{T}{\bf A}^m{\bf \hat W}{\bf s}_\text{E}^m - \sum_{m=1}^{N} \left ( {{\bm{\delta}}^m} \right )^\text{T}{\bf B}^m{\bf \hat W}{\bf c}_\text{E}^m \\
& = \min_{\left\{{{\bm{\delta}}^m}\right\}, {\mu}} {\kern 1pt} \frac{1}{2 \mu} \sum_{m=1}^{N} \sum_{n=1}^{N} \left ( {{\bm{ \delta}}^m} \right ) ^\text{T}{\bf A}^m \left ( {\bf A}^n \right )^\text{T}{\bm{\delta}}^n \left ( {{\bf s}_\text{E}^n} \right )^\text{T}{\bf D}^{-1} {{\bf s}_\text{E}^m} + \frac{1}{2 \mu} \sum_{m=1}^{N} \sum_{n=1}^{N} \left ( {{\bm{\delta}}^m} \right ) ^\text{T}{\bf A}^m \left ( {\bf B}^n \right )^\text{T}{\bm{\delta}}^n \left ( {{\bf c}_\text{E}^n} \right )^\text{T}{\bf D}^{-1} {{\bf s}_\text{E}^m}   \\
& {\kern 36pt} + \frac{1}{2 \mu} \sum_{m=1}^{N} \sum_{n=1}^{N} \left ( {{\bm{\delta}}^m} \right ) ^\text{T}{\bf B}^m \left ( {\bf A}^n \right )^\text{T}{\bm{\delta}}^n \left ( {{\bf s}_\text{E}^n} \right )^\text{T}{\bf D}^{-1} {{\bf c}_\text{E}^m} + \frac{1}{2 \mu} \sum_{m=1}^{N} \sum_{n=1}^{N} \left ( {{\bm{\delta}}^m} \right ) ^\text{T}{\bf B}^m \left ( {\bf B}^n \right )^\text{T}{\bm{\delta}}^n \left ( {{\bf c}_\text{E}^n} \right )^\text{T}{\bf D}^{-1} {{\bf c}_\text{E}^m}\\
&= \min_{\left\{{{\bm{\delta}}^m}\right\}, {\mu}} \frac{1}{2 \mu} \sum_{m=1}^{N} \sum_{n=1}^{N} \left ( {{\bm{\delta}}^m} \right ) ^\text{T}\left [ p_{m,n} {\bf A}^m \left ( {\bf A}^n \right )^\text{T} + f_{m,n} {\bf A}^m \left ( {\bf B}^n \right )^\text{T} + g_{m,n} {\bf B}^m \left ( {\bf A}^n \right )^\text{T} + q_{m,n} {\bf B}^m \left ( {\bf B}^n \right )^\text{T} \right ] {\bm{\delta}}^n
\end{aligned}
\label{eq_22}
\end{equation}

\hrulefill
\vspace*{4pt}
\end{figure*}

$\mathcal{P}_0$ is a joint optimization over all symbol slots, and it is a convex problem that can be directly solved via optimization tools such as CVX. To facilitate subsequent derivations, we introduce $\bf \hat W$:
\setcounter{equation}{6}
\begin{equation}
{\bf \hat W} = \begin{bmatrix}
 \Re \left ( \bf W \right ) & -\Im \left ( \bf W \right )
\end{bmatrix} \in {\mathbb R}^{N_\text{T} \times 2K},
\label{eq_7}
\end{equation}
based on which we can decompose ${\bf W}_\text{E}$ into
\begin{equation}
{\bf W}_\text{E}  = {\bf P} {\bf \hat W} + {\bf Q} {\bf \hat W} {\bf T},
\label{eq_8}
\end{equation}
where ${\bf P} \in {\mathbb R}^{2N_\text{T} \times N_\text{T}}$, ${\bf Q} \in {\mathbb R}^{2N_\text{T} \times N_\text{T}}$ and ${\bf T} \in {\mathbb R}^{2K \times 2K}$ are defined as
\begin{equation}
{\bf P}= \left [ \begin{matrix}
{\bf I}_{N_\text{T}}  \\
{\bf 0}
\end{matrix} \right ], {\kern 3pt} {\bf Q}=\left [ \begin{matrix}
{\bf 0}  \\
{\bf I}_{N_\text{T}}
\end{matrix} \right ], {\kern 3pt} {\bf T}= \left [ \begin{matrix}
 {\bf 0} & {\bf I}_K \\
 -{\bf I}_K & {\bf 0}
\end{matrix} \right ].
\label{eq_9}
\end{equation}
Based on \eqref{eq_8}, the expression for ${\bm \alpha}_\text{E}^n$ is further transformed into:
\begin{equation}
\begin{aligned}
{\bm \alpha}_\text{E}^n &= {\bf M}^n {\bf W}_\text{E} {\bf s}_\text{E}^n \\
&={\bf M}^n \left( {{\bf P} {\bf \hat W} + {\bf Q} {\bf \hat W} {\bf T}} \right) {\bf s}_\text{E}^n \\
&={\bf M}^n {\bf P} {\bf \hat W} {\bf s}_\text{E}^n + {\bf M}^n {\bf Q} {\bf \hat W} {\bf T} {\bf s}_\text{E}^n\\
&={\bf A}^n{\bf \hat W} {\bf s}_\text{E}^n + {\bf B}^n {\bf \hat W} {\bf c}_\text{E}^n,
\end{aligned}
\label{eq_10}
\end{equation}
where we introduce ${\bf A}^n \in {\mathbb R}^{2K \times N_\text{T}}$, ${\bf B}^n \in {\mathbb R}^{2K \times N_\text{T}}$ and ${\bf c}_\text{E}^n \in {\mathbb R}^{2K \times 1}$ as
\begin{equation}
{\bf A}^n= {\bf M}^n {\bf P}, {\kern 3pt} {\bf B}^n={\bf M}^n {\bf Q}, {\kern 3pt} {\bf c}_\text{E}^n={\bf T} {\bf s}_\text{E}^n.
\label{eq_11}
\end{equation}
With the expression for the $k$-th entry of ${\bm \alpha}_\text{E}^n$ given by
\begin{equation}
\alpha_k^n = \left( {{\bf a}_k^n} \right)^\text{T}{\bf \hat W} {\bf s}_\text{E}^n + \left( {{\bf b}_k^n} \right)^\text{T}{\bf \hat W} {\bf c}_\text{E}^n,
\label{eq_12}
\end{equation}
$\mathcal{P}_0$ can be expressed in the form of a standard convex optimization problem below:
\begin{equation}
\begin{aligned}
&\mathcal{P}_1: {\kern 3pt} \min_{{\bf \hat W}, {t}} - {t} \\
&\text{s.t.} {\kern 1pt} {\bf C1:} {\kern 2pt}  {t} - \left( {{\bf a}_k^n} \right)^\text{T}{\bf \hat W} {\bf s}_\text{E}^n - \left( {{\bf b}_k^n} \right)^\text{T}{\bf \hat W} {\bf c}_\text{E}^n \le 0, {\kern 1pt} \forall k \le 2K, n \le N, \\
&{\kern 12.5pt} {\bf C2:} {\kern 2pt}  \sum_{n=1}^{N} \left \| {\left({{\bf P} {\bf \hat W} + {\bf Q} {\bf \hat W} {\bf T}}\right) {\bf s}_\text{E}^n  }  \right \| _2^2 - N{p_0} \le 0.
\label{eq_13}
\end{aligned}
\end{equation}

\subsection{Closed-Form Structure for $\bf \hat W$}
We analyze $\mathcal{P}_1$ based on the Lagrangian and KKT conditions to derive the optimal precoding matrix $\bf \hat W$. The Lagrangian of $\mathcal{P}_1$ can be constructed as shown in \eqref{eq_14} on the top of this page, where ${\bm{\delta}}^n = \left[ \delta_1^n, \delta_2^n, \cdots, \delta_{2K}^n \right]^\text{T} \in {\mathbb R}^{2K \times 1}$ and $\mu$ are the non-negative dual variables associated with the inequality constraints ${\bf C1}$ and ${\bf C2}$ respectively, where we note that
\begin{equation}
\setcounter{equation}{15}
{\bf P}^\text{T}{\bf P}={\bf Q}^\text{T}{\bf Q} = {\bf I}_{N_\text{T}}, {\kern 3pt} {\bf P}^\text{T}{\bf Q} = {\bf Q}^\text{T}{\bf P} = {\bf 0}.
\label{eq_15}
\end{equation}
Accordingly, the KKT conditions for the optimality of $\mathcal{P}_1$ can be formulated and are shown in (16) on the top of next page. Based on the KKT conditions, we first obtain that $\mu >0$, otherwise ${\bm \delta}^n={\bf 0}$, $\forall n$, which contradicts with \eqref{eq_16a}. This means that the power constraint is active when the optimality is achieved, i.e.,
\begin{equation}
\setcounter{equation}{17}
 \sum_{n=1}^{N} \left ( {\bf s}_\text{E}^n  \right )^\text{T} {\bf \hat W}^\text{T} {\bf \hat W} {\bf s}_\text{E}^n + \sum_{n=1}^{N} \left ( {\bf c}_\text{E}^n  \right )^\text{T} {\bf \hat W}^\text{T} {\bf \hat W} {\bf c}_\text{E}^n = Np_0.
\label{eq_17}
\end{equation}
To proceed, we transform \eqref{eq_16b} into
\begin{equation}
2{\mu} {\bf \hat W} {\bf D} = \sum_{n=1}^{N} \left [ {\left ( {\bf A}^n \right ) }^\text{T} {{\bm{\delta}}^n} \left ( {\bf s}_\text{E}^n  \right )^\text{T} + {\left ( {\bf B}^n \right ) }^\text{T} {{\bm{\delta}}^n} \left ( {\bf c}_\text{E}^n  \right )^\text{T} \right ],
\label{eq_18}
\end{equation}
where ${\bf D} \in {\mathbb R}^{2K \times 2K}$ is given by
\begin{equation}
{\bf D} = \left [ {\sum_{n=1}^{N} {\bf s}_\text{E}^n  \left ( {\bf s}_\text{E}^n \right )^\text{T} + \sum_{n=1}^{N} {\bf c}_\text{E}^n  \left ( {\bf c}_\text{E}^n \right )^\text{T} }  \right ].
\label{eq_19}
\end{equation}
Based on the fact that the block length $N$ is in general larger than the number of users $K$, $\bf D$ is thus full-rank and invertible \cite{3GPP}. Accordingly, we can obtain an expression for the optimal precoding matrix ${\bf \hat W}$ as a function of the Lagrange multipliers ${\bm \delta}^n$ in a closed form as
\begin{equation}
{\bf \hat W} = \frac{1}{2{\mu}} \sum_{n=1}^{N} \left [ {\left ( {\bf A}^n \right ) }^\text{T} {{\bm{\delta}}^n} \left ( {\bf s}_\text{E}^n  \right )^\text{T} + {\left ( {\bf B}^n \right ) }^\text{T} {{\bm{\delta}}^n} \left ( {\bf c}_\text{E}^n  \right )^\text{T} \right ] {\bf D}^{-1}.
\label{eq_20}
\end{equation}
We observe from \eqref{eq_20} that the expression for the optimal precoding matrix $\bf \hat W$ includes all the data symbols transmitted within the block. In what follows, we consider the dual problem formulation of $\mathcal{P}_1$ to further simplify the CI-BLP problem.

\subsection{Dual Problem Formulation}
It is obvious that the Slater's condition is satisfied for the convex CI-BLP optimization problem $\mathcal{P}_1$ in \eqref{eq_13} \cite{r35}. Accordingly, we can solve $\mathcal{P}_1$ optimally by solving its dual problem, as shown below
\begin{equation}
\setcounter{equation}{21}
{\cal U}= \max_{\left\{{{\bm{\delta}}^m}\right\}, {\mu}} \min_{{\bf \hat W}, {t}}  {\cal L}\left ( {\bf \hat W} , {t}, {{\bm{\delta}}^m} , {\mu}  \right ),
\label{eq_21}
\end{equation}
where the inner minimization is achieved with \eqref{eq_16a}, the active power constraint in (17) and ${\bf \hat W}$ in \eqref{eq_20}. By substituting \eqref{eq_16a}, (17) and \eqref{eq_20} into ${\cal U}$ in (21), and by defining
\begin{equation}
\begin{aligned}
\setcounter{equation}{23}
p_{m,n}&=\left ( {{\bf s}_\text{E}^n} \right )^\text{T}{\bf C}^{-1} {{\bf s}_\text{E}^m}, {\kern 3pt} q_{m,n}= \left ( {{\bf c}_\text{E}^n} \right )^\text{T}{\bf C}^{-1} {{\bf c}_\text{E}^m}, \\
f_{m,n}&=\left ( {{\bf c}_\text{E}^n} \right )^\text{T}{\bf C}^{-1} {{\bf s}_\text{E}^m}, {\kern 3pt} g_{m,n}= \left ( {{\bf s}_\text{E}^n} \right )^\text{T}{\bf C}^{-1} {{\bf c}_\text{E}^m},
\end{aligned}
\label{eq_23}
\end{equation}
the objective function of the dual problem ${\cal U}$ can be simplified and is given by \eqref{eq_22} above. Further defining
\begin{equation}
\setcounter{equation}{24}
{{\bm{\delta}}_\text{E}} = \left[ {\left ( {{\bm{\delta}}^1} \right ) ^\text{T}, \left ( {{\bm{\delta}}^2} \right ) ^\text{T}, \cdots, \left ( {{\bm{\delta}}^N} \right ) ^\text{T}} \right]^\text{T} \in {\mathbb R}^{2NK \times 1}
\label{eq_25}
\end{equation}
and ${\bf U}_{m,n} \in {\mathbb R}^{2K \times 2K}$ given by
\begin{equation}
\begin{aligned}
{\bf U}_{m,n}=&{\kern 3pt} p_{m,n} {\bf A}^m \left ( {\bf A}^n \right )^\text{T} + f_{m,n} {\bf A}^m \left ( {\bf B}^n \right )^\text{T} + g_{m,n} {\bf B}^m \left ( {\bf A}^n \right )^\text{T} \\
&+ q_{m,n} {\bf B}^m \left ( {\bf B}^n \right )^\text{T},
\end{aligned}
\label{eq_26}
\end{equation}
the objective function of the dual problem ${\cal U}$ can finally be expressed as
\begin{equation}
{\cal U}=\min_{\left\{{{\bm{\delta}}^m}\right\}, {\mu}} \frac{1}{2 \mu} \left( {{{\bm{\delta}}_\text{E}}} \right)^\text{T} {\bf U} {{\bm{\delta}}_\text{E}},
\label{eq_27}
\end{equation}
where ${\bf U} \in {\mathbb R}^{2NK \times 2NK}$ is a block matrix constructed as
\begin{equation}
{\bf U} = \begin{bmatrix}
 {\bf U}_{1,1} & \cdots & \cdots& \cdots & {\bf U}_{1,N}\\
 \vdots & \ddots & {\bf U}_{m,n} & \ddots & \vdots \\
 {\bf U}_{N,1} & \cdots & \cdots & \cdots & {\bf U}_{N,N}
\end{bmatrix}.
\label{eq_28}
\end{equation}

\newcounter{mytempeqncnt5}
\begin{figure*}[!t]
\normalsize
\setcounter{equation}{27}

\begin{equation}
\begin{aligned}
&\sum_{l=1}^{N} \left ( {\bf s}_\text{E}^l  \right )^\text{T} {\bf \hat W}^\text{T} {\bf \hat W} {\bf s}_\text{E}^l \\
=& \frac{1}{4{\mu^2}} \sum_{l=1}^{N} \left ( {\bf s}_\text{E}^l  \right )^\text{T} \left \{ \sum_{m=1}^{N} \left [ \left ( {\bf A}^m \right )^\text{T}{\bm{\delta}}^m \left ( {\bf s}_\text{E}^m \right )^\text{T} + \left ( {\bf B}^m \right )^\text{T}{\bm{\delta}}^m \left ( {\bf c}_\text{E}^m \right )^\text{T} \right ] {\bf D}^{-1}  \right \}^\text{T} \left \{ \sum_{n=1}^{N} \left [ \left ( {\bf A}^n \right )^\text{T}{\bm{ \delta}}^n \left ( {\bf s}_\text{E}^n \right )^\text{T} + \left ( {\bf B}^n \right )^\text{T}{\bm{\delta}}^n \left ( {\bf c}_\text{E}^n \right )^\text{T} \right ] {\bf D}^{-1}  \right \} {\bf s}_\text{E}^l \\
=& \frac{1}{4{\mu^2}}\sum_{l=1}^{N}\sum_{m=1}^{N}\sum_{n=1}^{N} \left\{ \left ( {\bm{ \delta}}^m \right )^\text{T} \left [ {p_{l,n}} {p_{m,l}} {\bf A}^m \left ( {\bf A}^n \right )^\text{T} + {f_{l,n}} {p_{m,l}} {\bf A}^m \left ( {\bf B}^n \right )^\text{T} + {p_{l,n}} {g_{m,l}} {\bf B}^m \left ( {\bf A}^n \right )^\text{T} + {f_{l,n}} {g_{m,l}} {\bf B}^m \left ( {\bf B}^n \right )^\text{T} \right ] {\bm{\delta}}^n  \right\}
\end{aligned}
\label{eq_29}
\end{equation}

\hrulefill
\vspace*{4pt}
\end{figure*}

In the following, we simplify the block-level power constraint in (17). By substituting the expression for $\bf \hat W$ in \eqref{eq_20} into (17), the first term on the left-hand side of (17) is expanded and shown in \eqref{eq_29} on the top of next page. Based on the result in \eqref{eq_29} and by introducing ${\bf F}_{m,n}^l \in {\mathbb R}^{2K \times 2K}$ as
\begin{equation}
\begin{aligned}
\setcounter{equation}{29}
{\bf F}_{m,n}^l =& {\kern 3pt} {p_{l,n}} {p_{m,l}} {\bf A}^m \left ( {\bf A}^n \right )^\text{T} + {f_{l,n}} {p_{m,l}} {\bf A}^m \left ( {\bf B}^n \right )^\text{T} \\
&+ {p_{l,n}} {g_{m,l}} {\bf B}^m \left ( {\bf A}^n \right )^\text{T} + {f_{l,n}} {g_{m,l}} {\bf B}^m \left ( {\bf B}^n \right )^\text{T},
\label{eq_30}
\end{aligned}
\end{equation}
the first term on the left-hand side of (17) can finally be expressed in a compact form as
\begin{equation}
\sum_{l=1}^{N} \left ( {\bf s}_\text{E}^l  \right )^\text{T} {\bf \hat W}^\text{T} {\bf \hat W} {\bf s}_\text{E}^l = \frac{1}{4{\mu^2}} \left( {{{\bm{\delta}}_\text{E}}} \right)^\text{T} {\bf F}  {{\bm{ \delta}}_\text{E}},
\label{eq_31}
\end{equation}
where ${\bf F}= \sum_{l=1}^{N} {\bf F}^l \in {\mathbb R}^{2NK \times 2NK}$, with each ${\bf F}^l$ being a block matrix constructed with ${\bf F}_{m,n}^l$ similarly to the construction of ${\bf U}$ in \eqref{eq_28}.

Following the above procedure, the second term on the left-hand side of (17) can be similarly expressed in a compact form as:
\begin{equation}
\sum_{n=1}^{N} \left ( {\bf c}_\text{E}^n  \right )^\text{T} {\bf \hat W}^\text{T} {\bf \hat W} {\bf c}_\text{E}^n = \frac{1}{4{\mu^2}} \left( {{{\bm{\delta}}_\text{E}}} \right)^\text{T} {\bf G}  {{\bm{ \delta}}_\text{E}},
\label{eq_33}
\end{equation}
where ${\bf G}= \sum_{l=1}^{N} {\bf G}^l$. Each ${\bf G}^l \in {\mathbb R}^{2NK \times 2NK}$, $\forall l \le N$ is formulated with ${\bf G}_{m,n}^l$ similarly to the construction of ${\bf U}$ in \eqref{eq_28}, where ${\bf G}_{m,n}^l \in {\mathbb R}^{2K \times 2K}$ is given by
\begin{equation}
\begin{aligned}
{\bf G}_{m,n}^l = & {\kern 3pt} {g_{l,n}} {f_{m,l}} {\bf A}^m \left ( {\bf A}^n \right )^\text{T} + {q_{l,n}} {f_{m,l}} {\bf A}^m \left ( {\bf B}^n \right )^\text{T} \\
&+ {g_{l,n}} {q_{m,l}} {\bf B}^m \left ( {\bf A}^n \right )^\text{T} + {q_{l,n}} {q_{m,l}} {\bf B}^m \left ( {\bf B}^n \right )^\text{T}.
\end{aligned}
\label{eq_34}
\end{equation}
Based on the above derivations, the block-level power constraint (17) is equivalent to:
\begin{equation}
\begin{aligned}
&\frac{1}{4{\mu^2}} \left( {{{\bm{\delta}}_\text{E}}} \right)^\text{T} {\bf F}  {{\bm{ \delta}}_\text{E}} + \frac{1}{4{\mu^2}} \left( {{{\bm{\delta}}_\text{E}}} \right)^\text{T} {\bf G}  {{\bm{\delta}}_\text{E}} = Np_0 \\
\Rightarrow & \frac{1}{4{\mu^2}} \left( {{{\bm{\delta}}_\text{E}}} \right)^\text{T} \left({{\bf F}+{\bf G}}\right)  {{\bm{\delta}}_\text{E}} = Np_0.
\label{eq_35}
\end{aligned}
\end{equation}

By studying the relationship between $\left( {{\bf F} + {\bf G}} \right)$ and $\bf U$, we arrive at the following proposition.

{\bf Proposition 1:} $\bf F$, $\bf G$, and $\bf U$ satisfy the following condition:
\begin{equation}
{\bf F} + {\bf G} = {\bf U}.
\label{eq_37}
\end{equation}

{\bf Proof:} This proposition can be proved by expressing the generic $(m,n)$-th block in $\left( {\bf F} + {\bf G} \right)$ and $\bf U$, and through some mathematical transformations we can obtain ${\bf F}_{m,n} + {\bf G}_{m,n} = {\bf U}_{m,n}$, which completes the proof. $\blacksquare$

According to {\bf Proposition 1} and based on the block-level power constraint in \eqref{eq_35}, we can obtain the following expression for $\mu$:
\begin{equation}
\frac{1}{4{\mu^2}} \left( {{{\bm{\delta}}_\text{E}}} \right)^\text{T} {\bf U}  {{\bm{\delta}}_\text{E}} = Np_0 {\kern 3pt}
\Rightarrow {\kern 3pt}  {\mu} = \sqrt{\frac{\left( {{{\bm{\delta}}_\text{E}}} \right)^\text{T} {\bf U}  {{\bm{\delta}}_\text{E}}}{4Np_0}}.
\label{eq_38}
\end{equation}
Substituting the above expression for $u_0$ into \eqref{eq_27}, ${\cal U}$ can be transformed into an optimization on ${{\bm{\delta}}_\text{E}}$ only, given by
\begin{equation}
\begin{aligned}
{\cal U} &= \min_{{\bm{\delta}}_\text{E}, {\mu}} \frac{1}{2{\mu}} {\bm {\delta}}_\text{E}^\text{T} {\bf U} {\bm {\delta}}_\text{E}\\
& = \min_{{\bm{\delta}}_\text{E}} \frac{1}{2\sqrt{\frac{\left( {{{\bm{\delta}}_\text{E}}} \right)^\text{T} {\bf U}  {{\bm{\delta}}_\text{E}}}{4Np_0}}} {\bm {\delta}}_\text{E}^\text{T} {\bf U} {\bm {\delta}}_\text{E}\\
& = \min_{{\bm{\delta}}_\text{E}} \sqrt{N p_0 {\bm {\delta}}_\text{E}^\text{T} {\bf U} {\bm { \delta}}_\text{E}} \\
& = \min_{{\bm{\delta}}_\text{E}} {\bm {\delta}}_\text{E}^\text{T} {\bf U} {\bm {\delta}}_\text{E},
\label{eq_39}
\end{aligned}
\end{equation}
where the last step is achieved because $y=\sqrt{x}$ is a monotonic function. Accordingly, the final dual problem of the proposed CI-BLP optimization for PSK modulation can be formulated as
\begin{equation}
\begin{aligned}
&\mathcal{P}_2: {\kern 3pt} \min_{{\bm{\delta}}_\text{E}} {\bm {\delta}}_\text{E}^\text{T} {\bf U} {\bm {\delta}}_\text{E} \\
&{\kern 6pt} \text{s.t.} {\kern 11pt} {\bf C1:}{\kern 3pt} {\bf 1}^\text{T} {\bm {\delta}}_\text{E} -1=0 , \\
&{\kern 29pt} {\bf C2:}{\kern 3pt} \delta_\text{E}^m \ge 0, {\kern 3pt} \forall m \in \left\{ {1,2,\cdots, 2NK} \right\},
\label{eq_40}
\end{aligned}
\end{equation}

$\mathcal{P}_2$ is a QP optimization problem over a simplex, which can be more efficiently solved than the original CI-BLP optimization problem $\mathcal{P}_1$ via the standard simplex method \cite{r36}, \cite{r37} or the interior-point methods \cite{r38}. After solving $\mathcal{P}_2$ and obtaining ${\bf \hat W}$ via \eqref{eq_20}, the original complex precoding matrix $\bf W$ in \eqref{eq_1} can be obtained by
\begin{equation}
{\bf W}= {\bf \hat W} {\bf \hat P} - \jmath {\bf \hat W} {\bf \hat  Q},
\label{eq_41}
\end{equation}
where the form of ${\bf \hat P}$ and ${\bf \hat Q}$ follows \eqref{eq_9} while their dimension is changed into $2K \times K$.

\section{Numerical Results}
We present numerical results to validate our above derivations and illustrate the superiority of the proposed CI-BLP approach. Throughout the simulations, $K=N_\text{T}=12$, and 8PSK modulation is employed. The transmit power budget per symbol slot is set as $p_0=1$, leading to the total transmit power budget for the considered block of symbol slots as $P_\text{total}=N p_0 =N$. We compare CI-BLP with block-level ZF-based precoding and traditional CI-SLP under standard Rayleigh fading channels.

\begin{figure*}[!t]
\captionsetup[subfigure]{labelformat=empty}
\captionsetup{labelformat=empty}
\begin{centering}
\subfloat[Fig. 2  SER v.s. SNR, $N=15$]
{\begin{centering}
\includegraphics[width=0.3\textwidth]{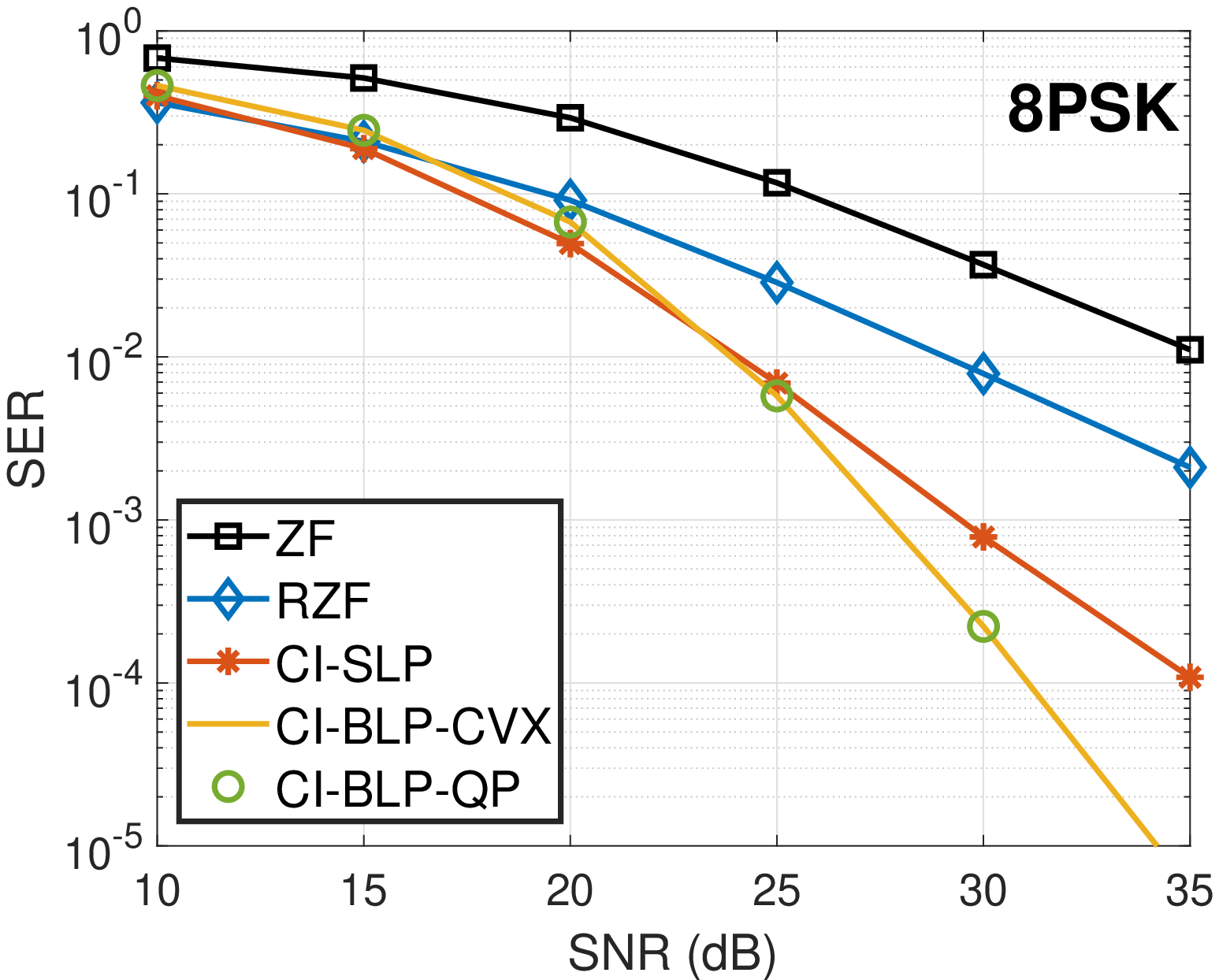}
\par
\end{centering}
}
\hspace{0.1cm}
\subfloat[Fig. 3  SER v.s. block length $N$]
{\begin{centering}
\includegraphics[width=0.3\textwidth]{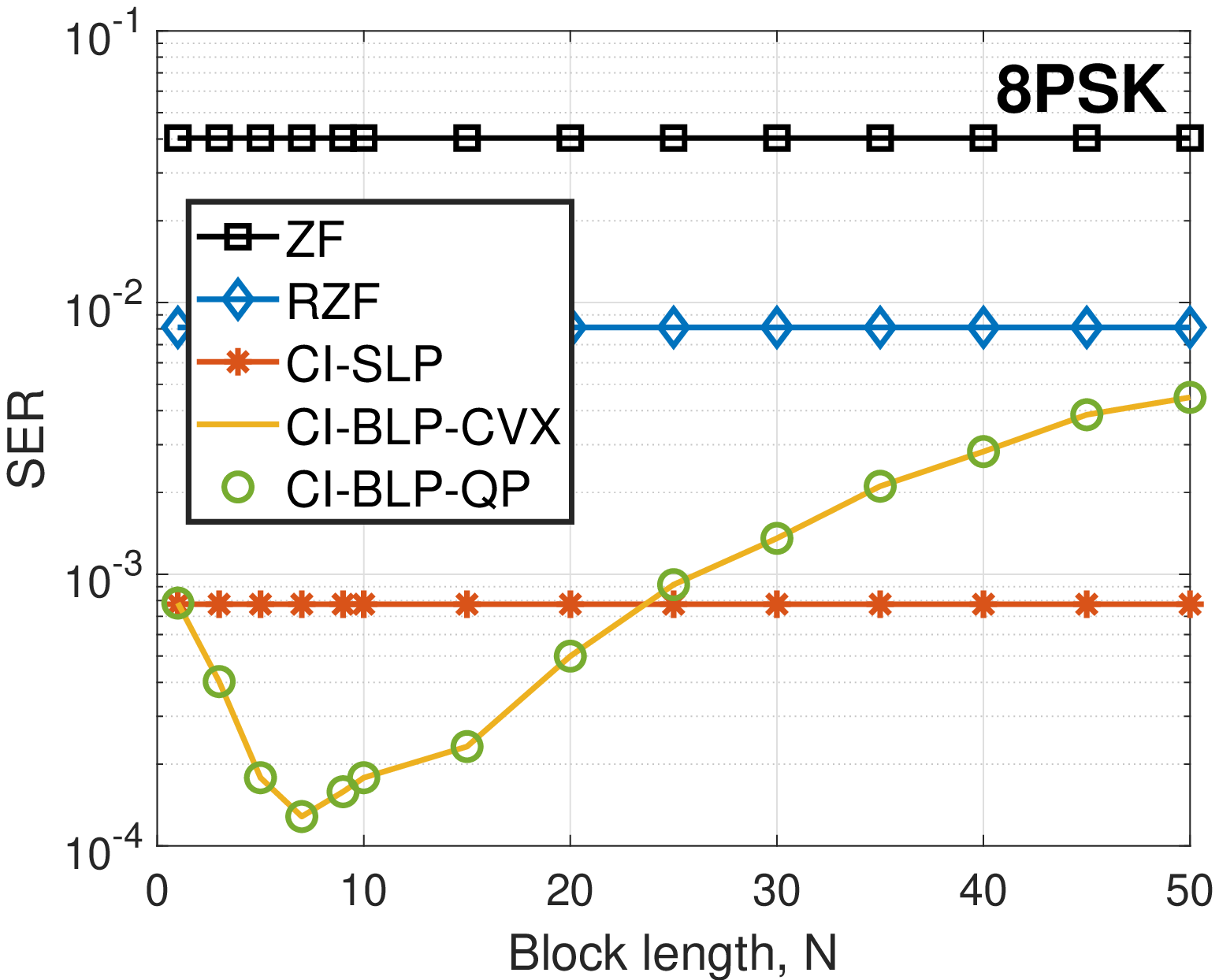}
\par
\end{centering}
}
\hspace{0.1cm}
\subfloat[Fig. 4  Execution time v.s. block length $N$]
{\begin{centering}
\includegraphics[width=0.3\textwidth]{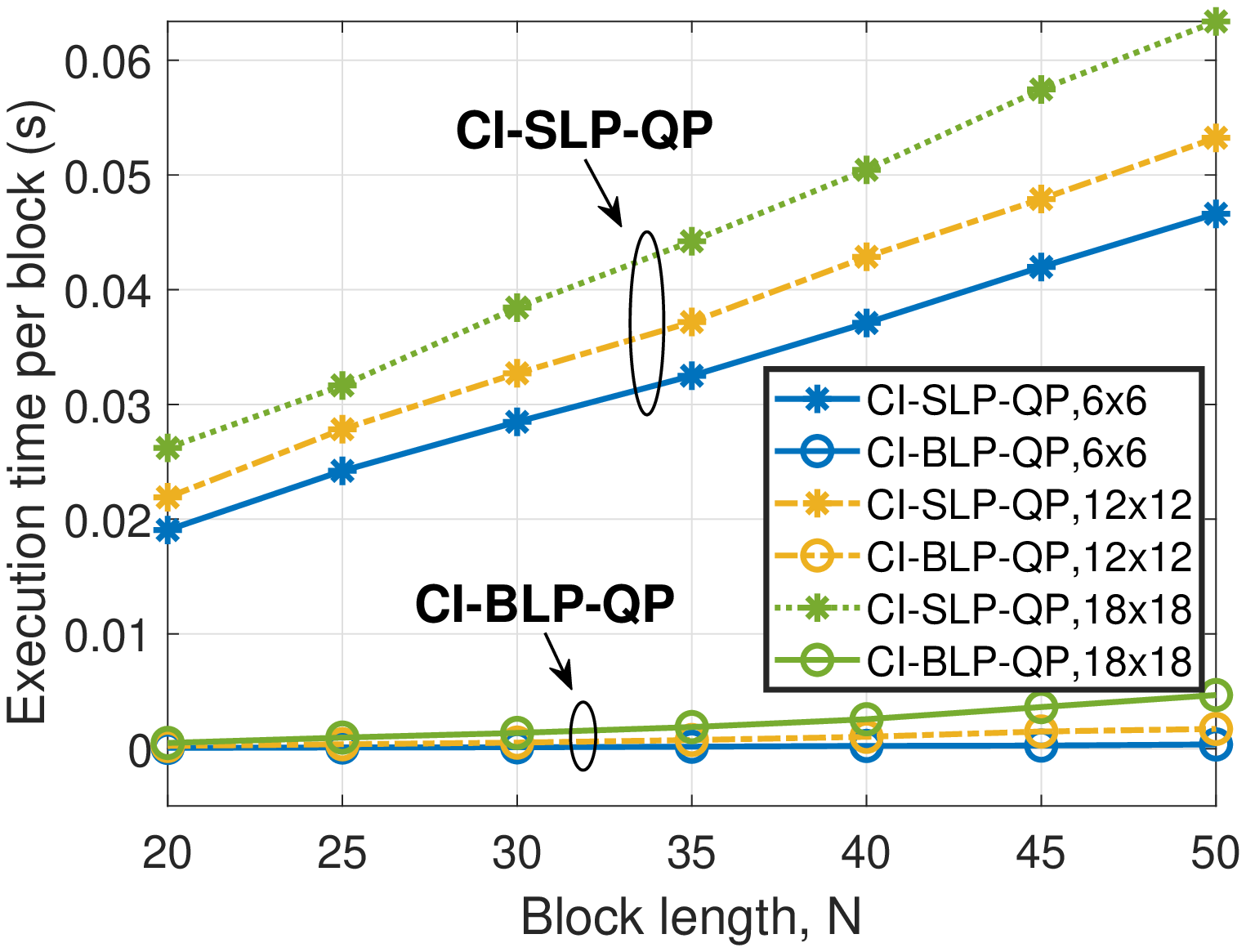}
\par
\end{centering}
}
\par
\end{centering}
\caption{Numerical comparisons for the proposed CI-BLP approach, 8PSK, $K=N_\text{T}=12$}
\end{figure*}

Fig. 3 depicts the symbol-error rate (SER) performance of different BLP and SLP approaches, where the block length is $N=15$. Compared with BLP approaches (ZF/RZF), CI-based precoding methods achieve an improved SER performance by exploiting CI. Thanks to the relaxed block-level power constraint, the CI-BLP method proposed in this paper offers additional performance improvements over traditional CI-SLP methods in the literature with reduced computational costs, making CI-BLP more attractive in practical MIMO communication systems. The results in Fig. 3 also validate the correctness of our derivations in the paper.

Fig. 4 presents the SER performance of the proposed CI-BLP scheme with respect to the block length $N$, where the transmit SNR is fixed at 30dB. The block length $N$ does not affect the design of ZF and RZF precoding, and therefore their SER remains constant. Interestingly, as the block length $N$ increases, CI-BLP's SER performance first improves thanks to the relaxed power constraint, which outweighs the loss due to using a constant precoding matrix over the block. As $N$ further increases, the SER performance becomes worse because relaxed power constraint cannot further compensate for the loss of using a constant precoding matrix.

Fig. 5 evaluates the computational complexity gain of the proposed CI-BLP method over traditional CI-SLP in terms of the execution time, where results for $6 \times 6$, $12 \times 12$ and $18 \times 18$ MU-MISO systems are presented. For fairness of comparison, only the execution time required for running the `quadprog' function in MATLAB used to solve the QP problem for both CI-BLP and CI-SLP is evaluated. We observe from Fig. 5 that the proposed CI-BLP approach offers a significant complexity gain over traditional CI-SLP, and the complexity gains become more prominent as the block length $N$ increases.

\section{Conclusions}
A block-level interference exploitation precoding termed CI-BLP is proposed for a downlink MU-MISO communication system, where a constant precoding matrix is applied to a block of data symbols, thus removing the symbol-by-symbol optimization required in traditional SLP and reducing the computational costs of CI-based precoding. The proposed CI-BLP optimization problem is shown to be equivalent to a QP problem over a simplex. Thanks to the relaxed block-level power constraint, a superior performance for the proposed CI-BLP scheme over traditional CI-SLP precoding is observed when the length of the symbol slots is short, while only a slight performance loss is exhibited as the length of the symbol slots increases, as validated by the numerical results.

\bibliographystyle{IEEEtran}
\bibliography{refs.bib}

\end{document}